\documentclass[12pt,reqno]{article}
\usepackage{amsmath}
\usepackage{graphicx}
\usepackage{color}
\usepackage{varioref}

\begin{document}

\title
\centerline{\Large{\bf{Shadowing corrections to the evolution of}}}\\
\centerline{\Large{\bf{singlet structure function}}}

\vspace{20pt}

\centerline{{\bf Mayuri Devee}}
\centerline{{Department of Physics, University of Science and Technology Meghalaya, Ri-Bhoi, Meghalaya 793101, India}} 
\centerline{{\it deveemayuri@gmail.com}}

\begin{abstract}
We report a semi-analytical approach for the solution of nonlinear Gribov-Levin-Ryskin-Mueller-Qiu (GLR-MQ) evolution equation for sea quark distribution and investigate the effect of gluon shadowing on the small-$x$ and moderate-$Q^2$ behaviour of singlet structure function, ${F_2^S(x,Q^2)}$. The predicted values of singlet structure functions with shadowing corrections are comparable with different experimental data as well as parametrizations. 
It is very fascinating to observe signatures of gluon recombination in our predictions. We note that, with decreasing $x$ the the rapid growth of singlet structure function is eventually tamed by the shadowing or nonlinear term appeared in the QCD evolution due to gluon recombination. This taming behaviour of shadowing singlet structure function towards small-$x$ become significant at the hot spots, at $R=2$ GeV$^{-1}$. Further the computed values of the ratio of ${F_2^S(x,Q^2)}$ with shadowing to that without shadowing clearly indicate that towards smaller values of $x$ and $Q^2$ the nonlinear effects play an increasingly important role.



\end{abstract}

\section{Introduction}

\ Exploring the dynamics of the regime of high gluon density is one of the extremely demanding contemporary issues,
in the domain of high energy or small-$x$ physics, where $x$ is the Bjorken scaling variable.
The gluon saturation is one of the most fascinating problems of the small-$x$ physics, which is presumed on theoretical basis and there is emerging evidence of its existence [1-3].
The linear DGLAP evolution equations at the twist-$2$ level [4-6] speculates a sharp growth of the gluon densities towards small-$x$ which is observed in the deep inelastic scattering (DIS) experiments at HERA as well.
This generates cross sections which in the high-energy limit fail to comply with the unitarity bound or more specifically the  Froissart bound [7].
Subsequently the growing number of gluon densities, so as to approach small-$x$, demands a formulation of the QCD at high partonic density incorporating the unitarity corrections in a proper way.
Following DGLAP, the growing number of small-$x$ gluons graphically conforms to higher density of individuals in the same approved region, which differs from a more diluted system at moderate values of $x$.
Again, perturbative QCD manifests that the sea quark distributions, in a hadron evolves rapidly with decreasing $x$ at fixed $Q^2$ in the same manner as the gluon distribution $xg(x,Q^2)$.  However in the region of very small-$x$
the sharp growth of the sea quark density
is expected to slow down eventually in order to restore the Froissart bound on physical cross sections.
The sea quark distribution, which overshadows the valence quarks at small $x$, is supposed to be generated through gluons and therefore it is anticipated that the gluon and sea quark distribution functions may experience the same effect of shadowing.

\ In general the gluon recombination processes are considered to be responsible for this taming behaviour. At very small values of $x$ the probable interaction between two gluons can no longer be overlooked and it sooner or later
generates a situation in which individual partons as a matter of course overlap or shadow each other.
As a consequence, nonlinear phenomena are likely to occur which should eventually lead to a control of the maximum gluon density per unit of phase-space. The nonlinear or shadowing corrections in DIS arise due to two processes, one is the taming of the gluon density as a result of gluon recombination $gg\to{g}$ and the other is the Glauber-like rescattering of the $q\bar{q}$ fluctuations off gluons. The second process can also be regarded as a parton recombination, particularly as a recombination of gluons into a $q\bar{q}$ pair, $gg\to{q\bar{q}}$.
These multiple gluon interactions induce nonlinear corrections to the conventional linear DGLAP evolution equations. %

\ Gribov, Levin and Ryskin, at the onset, explored the shadowing corrections of gluon recombination to the parton distributions i.e quark and gluon distributions based on the Abramovsky-Gribov-Kancheli (AGK) cutting rules in the double leading logarithmic approximation (DLLA) [8]. Following that Mueller and Qiu completed the equation numerically using a perturbative calculation of the recombination probabilities in the DLLA, and also formulated the equation for the conversion of gluons to quarks [9, 10]. This is a triumph of great interest as it provides the connection to experiments and thus enables the equation to be applied phenomenologically. Their results are widely known as the GLR-MQ equation and can be regarded as the improved version of the conventional linear DGLAP equations with the corrections for gluon recombination.
The GLR-MQ equation incorporates the dominant non-ladder contributions, denoted as the fan diagrams, apart from the production diagrams in order to account for gluon recombination processes. The fan diagrams portrays the decisive role in the restoration of unitarity by taking into consideration some of the gluon recombination processes that become vital at small-$x$. The GLR-MQ equation predicts a critical line separating the perturbative regime from the saturation regime and it is legitimate only in the edge of this critical line.
Thus the study of the GLR-MQ equation may provide important insight into the nonlinear effects of gluon recombination due to the high gluon density
at sufficiently small $x$.

\ Most of the analysis on the modifications of the higher order QCD effects is established on the semi-classical [8, 11] as well as on numerical approach [12-15]. However, some analytical approaches are also reported in recent years [16-17] to study the nonlinear GLR-MQ evolution equation. It is always very alluring to explore the prospect of obtaining analytical solutions of GLR-MQ equation.
Recently we report the approximate analytical solution of the nonlinear GLR–MQ evolution equation for gluon distribution at leading order by using the Regge-like ansatz and investigate the effect of shadowing corrections to the small-$x$ and moderate $Q^2$-behaviour of gluon distribution function with considerable phenomenological success [18, 19]. Recently, the same approach has been employed in [20-22] to obtain the solution of GLR-MQ beyond leading order.
In the present paper, we intend to solve the GLR-MQ equation in a semi-analytical approach for sea-quark distribution, somewhat in the restricted domain of $x$ and $Q^2$, by incorporating the well-known Regge ansatz in the leading twist approximation. Here we obtain the small-$x$ and moderate-$Q^2$-behaviour of singlet structure function $F_2(x,Q^2)$, and examine the signatures of shadowing due to gluon recombination in our predictions. We further evaluate the analytical expressions for $F_2(x,Q^2)$ by solving the linear DGLAP equation using the Regge ansatz and
make a comparative analysis of the nonlinear GLR-MQ equation and the linear DGLAP equation to examine the effect of nonlinearity in our predictions.
Our computed values of singlet structure function, $F_2^S(x,Q^2)$, with shadowing corrections are compared with the CERN's NMC [23], Fermilab E665 Collaboration [24] as well as with those obtained in the NNPDF [25] collaboration. 

\ The outline of the paper is: in Section 2 we present the general formalism to solve the GLR-MQ equation for sea-quark distribution and obtain the small-$x$ and moderate-$Q^2$-behaviour of singlet structure function $F_2^S(x,Q^2)$. The comparative analysis of nonlinear GLR-MQ equation and linear DGLAP equation for singlet structure function is also given in section 2. The results and discussions of our predictions of $F_2(x,Q^2)$ invoking shadowing corrections are presented in section 3. We summarize and conclude in Section 4.

\section{Formalism}
\subsection{General framework}

\ Two processes in the parton cascade contribute to the GLR-MQ equation, namely, the gluon splitting generated by the QCD vertex $g{\rightarrow}g+g$ as well as the gluon recombination by the same vertex $g+g{\rightarrow}g$. The probability that a gluon splits into two gluons is proportional to $\alpha_s\rho$ whereas the probability of gluon recombination is proportional to $\alpha_{s}^2{r^2}{\rho}^2$ [26, 27]. Here, $r$ is the size of the gluon produced in the recombination process and $r{\propto}\frac{1}{Q}$ concerning DIS.
Thus the number of partons in a phase space cell ($\Delta{\ln(1/x)}\Delta{\ln{Q^2}}$) increases through gluon splitting and decreases through gluon recombination and subsequently the nonlinear corrections emerged from the recombination of two gluon ladders modify the evolution equations of sea quark distribution as [8, 9, 28]
\begin{equation}\frac{\partial{xq(x,Q^2)}}{\partial{\ln{Q^2}}}=\frac{\partial{xq(x,Q^2)}}{\partial{\ln{Q^2}}}\Big\vert_{DGLAP}-\frac{27}{160}\frac{\alpha_S^2(Q^2)}{R^2Q^2}[xg(x,Q^2)]^2+HT,\end{equation}
which is known as the GLR-MQ evolution equation for sea quark distribution.

\ The first term in the r.h.s. of Eq. (1) is the normal DGLAP term in the DLLA and and is therefore linear in the gluon field. The second term carries a negative sign and it reduces the growth of the gluon distribution once the fan diagrams become admissible, i.e., at small-$x$. It expresses the non-linearity in respect of the square of the gluon distribution. $q(x,Q^2)$ is the quark density and $g(x,Q^2)$ is the gluon density. Here, the representation for the gluon distribution $G(x,Q^2)=xg(x,Q^2)$ is used.
The quark-gluon emission diagrams are not given attention here due to their trivial importance in the gluon-rich small-$x$ region.
$HT$ denotes a higher-dimensional gluon distribution term revealed by Mueller and Qiu but it is not given in all respects [9]. Therefore this term is not taken into account in our analysis presented below. The quark-gluon emission diagrams are also not considered here as they have very little effect in the region of small-$x$ which is governed predominantly by the gluons. The region of validity of Eq. (1) is not fully known, but a general criterion is that the non-linear correction term should not be larger than the first term since in that case further corrections must be considered and non-perturbative effects could be of importance [9].
The size of the nonlinear term crucially depends on the value of the correlation radius $R$ between two interacting gluons. ${\pi}R^2$ is the target area populated by the gluons. If the gluons originate from sources which occupy distinct regions in longitudinal coordinate space then $R$ is of the order of proton radius, i.e. $R=5$ GeV$^{-1}$. In that case recombination probability is very negligible. On the other hand, if the gluon ladders couple to the same parton (quark or gluon) then it leads to a higher gluon density in the parton's vicinity. Such hot spots [29, 30] of high gluon density can enumerate the rapid onset of gluon-gluon interactions in the environs of the emitting parton and so uplift the recombination effect or shadowing corrections. In such hot spots where $R$ is considered to be of the order of the transverse size of a valence quark, i.e. $R=2$ GeV$^{-1}$.

\ In the parton model approximation, the structure functions are usually identified by summing quark distributions weighted by squared charges as usual
\begin{equation}F_2(x,Q^2)={\sum_i}e_i^2xq_i(x,Q^2) \end{equation}
where the sum implies summation over all flavours of quarks and anti-quarks and $e_i$ is the electric charge of a quark of type $i$. The $F_2$ structure functions measured in DIS can be written in terms of singlet and nonsinglet quark distribution functions as
\begin{equation}F_2=\frac{5}{18}F_2^S+\frac{3}{18}F_2^{NS}\end{equation}
It is well-known that at small-$x$ the nonsinglet contribution is negligible and can be ignored.
Therefore at small-$x$ Eq. (1) can be approximated as
\begin{equation}\frac{\partial{F_2^S(x,Q^2)}}{\partial{\ln{Q^2}}}=\frac{5}{18}\frac{\partial{F_2^S(x,Q^2)}}{\partial{\ln{Q^2}}}\Big\vert_{DGLAP}-\frac{5}{18}\frac{27}{160}\frac{\alpha_S^2(Q^2)}{R^2Q^2}G^2(x,Q^2),\end{equation}

\ The linear DGLAP equation for singlet structure function in the leading twist approximation is given by [31]
\begin{eqnarray}\frac{\partial{F_2^S(x,Q^2)}}{\partial{\ln{Q^2}}}\Big\vert_{DGLAP}&=&\frac{\alpha_s(Q^2)}{2\pi}\bigg[\frac{2}{3}\Big(3+4\ln(1-x)\Big)F_2^S(x,Q^2)\nonumber \\
&& +\: \frac{4}{3}\int_x^1\frac{{d\omega}}{1-\omega}\Big\{(1+\omega^2)F_2^S\Big(\frac{x}{\omega},Q^2\Big)-2F_2^S(x,Q^2)\Big\} \nonumber \\
&& +\: N_F\int_x^1\Big(\omega^2+(1-\omega)^2\Big)G\Big(\frac{x}{\omega},Q^2\Big)d\omega\bigg],\end{eqnarray}
where, in the leading twist approximation the running coupling constant $\alpha_S(Q^2)$ has the form
\begin{equation}\frac{\alpha_s(Q^2)}{2\pi}=\frac{2}{\beta_0\ln(Q^2/\Lambda^2)},\end{equation}
where ${{\beta_{0}}={11-{{\frac{2}{3}}{N_{f}}}}}$ is the one-loop correction to the QCD $\beta$-function, $N_f$ is the number of active quark flavours and $\Lambda$ is the QCD cut off parameter.

\subsection{Shadowing corrections to the singlet structure function at small-$x$}

We make an attempt to solve the nonlinear GLR-MQ equation for singlet quark distribution given by Eq. (4) at small-$x$, by employing the well-known Regge like behaviour of singlet structure function.
It is interesting to note that the small-$x$ behaviour of structure functions can be successfully described in the framework of Regge theory [32]. The Regge theory is a naive and frugal parameterization of all total cross sections and is supposed to be applicable at large-$Q^2$ values, so that that a perturbative treatment is possible, if $x$ is small enough $x<0.07$ [33, 34]. Again, as the structure functions are proportional to the total virtual photon-nucleon cross section therefore they are expected to have Regge behaviour corresponding to pomeron or reggeon exchange.
The high energy attitude of hadronic cross sections as well as structure functions will be governed by two contributions according to the Donnachie-Landshoff (DL) model, especially a pomeron, proliferating the rise of structure function at small-$x$ and reggeons related with meson trajectories [33, 34].
The Regge pole model gives the parametrization of the DIS structure function $F_2(x,Q^2)$ at small-$x$ as $F_2\propto{x^{-\lambda}}$ with $\lambda>0$ being a constant or depending on $Q^2$ or $x$ [32].

\ On that account we employ the Regee like ansatz of singlet structure function to solve Eq. (4).
It is always a convenient idea to try the simplest assumptions and so we assume a simple form of Regge ansatz for singlet structure function as
\begin{equation}F_2^S(x,Q^2)=H(Q^2)x^{-\lambda_S},\end{equation}
where $H(Q^2)$ is a function of $Q^2$ and $\lambda_S$ is the Regge intercept for singlet structure function.
The high energy i.e. small-$x$ behaviour of both gluons and sea quarks are controlled by the same singularity factor in the complex angular momentum plane [32] in accordance with Regge thery since the same power is expected for sea quarks and gluons. Moreover the values of Regge intercepts for all the spin-independent singlet, non-singlet and gluon structure functions should be close to 0.5 in quite a broad range of small-$x$ [35]. Therefore in our present analysis we take the value of $\lambda_S$ to be 0.5. This value of $\lambda_S$ is also close to the intercept of the hard pomeron, $\epsilon_h$, in the DL two pomerons exchange model where $\epsilon_h=0.418$. 

\ The $Q^2$-evolution of the proton structure function $F_2(x,Q^2)$ is related to the gluon distribution function $G(x,Q^2)$ in the proton and to the strong interaction coupling constant $\alpha_S$. 
Hence the direct relations between $F_2(x,Q^2)$ and $G(x,Q^2)$ are extremely important because using those relations the experimental values of $G(x,Q^2)$ can be extracted using the data on $F_2(x,Q^2)$. A plausible way of realizing this is through the following ansatz [36,37] 
\begin{equation}{G(x,Q^2)=K(x)F_2^S(x,Q^2)}.\end{equation}
The evolution equations of gluon distribution function and singlet structure function are in the same forms of derivative with respect to $Q^2$. Moreover in global PDF analyses to incorporate different high precision data, particularly in the MSTW08 parton set, the input singlet and gluon parameterizations are assumed to be functions of $x$ at fixed $Q^2$ [38]. Accordingly the above assumption is justifiable.
The function $K(x)$ to be determined from phenomenological analysis. The actual functional form of $K(x)$ can be determined by simultaneous solutions of coupled equations of singlet structure functions and gluon parton densities, nevertheless it is beyond the scope of this paper.
Here we have performed our analysis considering the function $K(x)$ as an arbitrary constant parameter $K$ for a particular range of $x$ and $Q^2$ in defining the relation between gluon and singlet structure functions as the simplest assumption.
But, we need to adjust its value for satisfactory description of different experiments. The best fit graphs are obtained by choosing an appropriate value of $K$ for a proper description of each experiment.

Using Eq. (8) the term $G(\frac{x}{\omega},Q^2)$ can be written as
\begin{equation}G\Big(\frac{x}{\omega},Q^2\Big)=K\Big(\frac{x}{\omega}\Big)F_2^S(x,Q^2).\end{equation}


\ Substituting Eqs. (8) and (9) and employing the Regge ansatz of Eq. (7) for singlet structure functions in Eq. (4) we arrive at
\begin{equation}\frac{\partial{F_2^S(x,Q^2)}}{\partial{Q^2}}=p_1(x)\frac{F_2^S(x,Q^2)}{\ln(Q^2/\Lambda^2)}-p_2(x)\frac{\big[F_2^S(x,Q^2)\big]^2}{Q^2\ln(Q^2/\Lambda^2)},\end{equation}
The explicit forms of the functions $p_1(x)$ and $p_2(x)$ are
\begin{eqnarray}p_1(x)&=&\frac{5}{9\beta_0}\Bigg[\frac{2}{3}\Big(3+4\ln(1-x)\Big)+\frac{4}{3}\int_x^1\frac{{d\omega}}{1-\omega}\Big(\{(1+\omega^2)\omega^{\lambda_S}-2\Big) \nonumber \\
&& +\: N_F\int_x^1\Big(\omega^2+(1-\omega)^2\Big)\omega^{\lambda_S}K\Big(\frac{x}{\omega}\Big)d\omega\Bigg],\end{eqnarray}
\begin{equation}p_2(x)=\frac{27}{36}\frac{\pi^2\big(K(x)\big)^2}{\beta_0^2R^2}.\end{equation}
Eq. (10) is a partial differential equation for the singlet structure function $F_2(x, Q^2)$ with respect to the variables $x$ and $Q^2$. This equation can be used to examine the $x$-evolution of singlet structur function apart from its conventional use in $Q^2$-evolution. Solving of Eq. (10) we get
\begin{equation}F_2^S(x,Q^2)=\frac{t^{p_{1}(x)}}{C+p_{2}(x)\int{t}^{p_{1}(x)-2}\exp{(-t)}d{t}},\end{equation}
which leads us to the solution for the singlet structure function with nonlinear or shadowing corrections. Here we consider the variable $t$, such that $t=\ln(\frac{Q^2}{\Lambda^2})$, in order to simplify our calculations. The constant $C$ is to be determined from initial boundary conditions.

\ Here we restrict our analysis in the kinematic domain $0.4<{Q^2}<{30}$ GeV$^2$ and $10^{-5}<{x}<10^{-1}$ where the solution suggested in Eq. (13) is found to be a valid solution of the nonlinear GLR-MQ equation for singlet structure function. In this kinematic domain the solution suggested in Eq. (13) agrees well with the Regge ansatz of Eq. (7) and satisfactorily describes the shadowing corrections to the singlet structure function. Nevertheless, we note that at very large $t$ or in other words at very large $Q^2$ ($t=\ln(Q^2/\Lambda^2)$) as well as at large $x$ ($x>10^{-2}$), gluon recombination play less of a role on QCD evolution and accordingly the solution given by Eq. (13) is not legitimate at large-$Q^2$ and large-$x$.

\ We now use the following two physically plausible boundary conditions to evaluate the $Q^2$ and $x$ dependence of singlet structure function from Eq. (13).
\begin{equation}F_2^S(x, Q^2)=F_2^S(x, Q_0^2)\end{equation}
for some lower value of $Q^2=Q_0^2$  and
\begin{equation}F_2^S(x, Q^2)=F_2^S(x_0, Q^2),\end{equation}
at some high $x=x_0$.
Applying these boundary conditions in Eq. (13) we obtain the $Q^2$ and $x$ dependence of singlet structure function as
\begin{equation}F_2^S(x, Q^2)=\frac{t^{p_1(x)}F_2^S(x,t_0)}{{t_0^{p_1(x)}}+p_2(x)\Big[\int{t^{p_1(x)-2}\exp{(-t)}}dt-\int{t_0}^{p_1(x)-2}\exp{(-t_0)}dt_0\Big]F_2^S(x,t_0)}.\end{equation}
and
\begin{equation}F_2^S(x,Q^2)=\frac{t^{p_1(x)}F_2^S(x_0,t)}{{t^{p_1(x_0)}}+\Big[p_2(x)\int{t^{p_1(x)-2}\exp{(-t)}}dt-p_2(x_0)\int{t^{p_1(x_0)-2}\exp{(-t)}}dt\Big]F_2^S(x_0,t)}\end{equation}
respectively. Thus from Eq. (16) 
we can easily compute the dependence of $F_2^S(x, Q^2)$ on $Q^2$ for a particular value of $x$ by choosing an appropriate input distribution at a given value of $Q_0^2$. The effect of nonlinear or shadowing corrections to the singlet structure functions for a set of $Q^2$ can also be studied from this equation.
On the other hand, Eq. (17) provides us the small-$x$ dependence of nonlinear singlet structure function for a particular value of $Q^2$ with a suitable input distribution at an initial value of $x=x_0$. Eq. (17) further helps us to examine the effect of shadowing corrections to the small-$x$ dependence of singlet structure functions.

\subsection{Comparative analysis of singlet structure functions with and without shadowing}

\ In the derivation of the linear DGLAP equation, the correlations among the initial gluons in the physical process of interaction and recombination of gluons at small-$x$  are not usually taken into account. Therefore the solution of the linear DGLAP equation given by Eq. (5) leads us to the singlet structure function without shadowing corrections. Accordingly the comparative analysis of the singlet structure function obtained from the nonlinear GLR-MQ equation, which incorporates shadowing corrections, with that obtained from the linear DGLAP approach assist us to estimate the effect of nonlinearity in our predictions.
Now employing the Regge ansatz of Eq. (7) and following the same procedure as before Eq. (5) can be solved as
\begin{equation}F_2^S(x,Q^2)=Dt^{p_1(x)},\end{equation}
where $D$ is a constant to be fixed by initial boundary condition and $t=\ln(\frac{Q^2}{\Lambda^2})$. The $x$ dependent function $p_1(x)$ is defined in Eq. (11). Eq. (18) provides the solution of the linear singlet structure function without shadowing corrections.

Applying the boundary condition
\begin{equation}f_{10}\equiv{F_2^S(x,Q_0^2)}=Dt_0^{p_1(x)}\end{equation}
at $Q^2=Q_0^2$, in Eq. (18) we obtain  
\begin{equation}F_2^S(x,Q^2)=f_{10}\Big(\frac{t}{t_0}\Big)^{p_1(x)}.\end{equation}
This leads us to the 
$Q^2$-evolution of linear singlet structure function for a fixed value of $x$ provided a suitable input distribution $f_{10}$ has been chosen from the initial boundary condition.

Again, defining
\begin{equation}f_{20}\equiv{F_2^S(x_0,Q^2)}=Dt^{p_1(x_0)}\end{equation}
at some initial higher value $x=x_0$ Eq. (18) can be expressed as
\begin{equation}F_2^S(x,Q^2)=f_{20}t^{p_1(x)-p_1(x_0)}.\end{equation}
Thus Eq. (22)  describes the small-$x$ behavior of linear singlet structure function for a particular value of $Q^2$ by choosing an appropriate input distribution $f_{20}$ from the initial boundary condition.
\ 
\ Now from the solutions of the nonlinear GLR-MQ equation and the linear DGLAP equation given by Eqs. (17) and (22) respectively, we calculate the ratio of the function $F_2^S(x,Q^2)$ as 
\begin{equation}{R_{F_2^S(x,Q^2)}}=\frac{{F_2^S}(x,Q^2)^{GLR-MQ}}{{F_2^S}(x,Q^2)^{DGLAP}},\end{equation}
as a function of variable $x$ for different values of $Q^2$. 
From this ratio we can investigate how the gluon recombination processes effect the linear QCD evolution. Moreover, by analysing this ratio we understand that the shadowing corrections tame the behavior of the singlet structure function towards small-$x$ leading to a restoration of the Froissart bound. The phenomenological analysis of Eq. (23) is presented in section 3.

\section{Result and discussion}

\ We solve the nonlinear GLR-MQ evolution equation by considering the Regge like ansatz of singlet structure function and examine the effects of nonlinear or shadowing corrections due to gluon recombination processes at small-$x$ to the behavior of singlet structure function $F_2^S(x,Q^2)$. The small-$x$ and moderate-$Q^2$ dependence of $F_2^S(x,Q^2)$ are also investigated for both at $R=2$ GeV$^{-1}$ and $R=5$ GeV$^{-1}$ from the predicted solution of the GLR-MQ equation. 
To compute the dependence of $F_2^S(x,Q^2)$ on $Q^2$ we take the input distributions from the data point corresponding to the lowest value of $Q^2$ for a particular range of $Q^2$ under study. Similarly the data point corresponding to the highest value of $x$ of a particular range of $x$ under consideration are taken as input distribution to determine the $x$ dependence of $F_2^S(x,Q^2)$.
\begin{figure}[!htb]
\centering
\includegraphics[width=2.6in, height=2.7in]{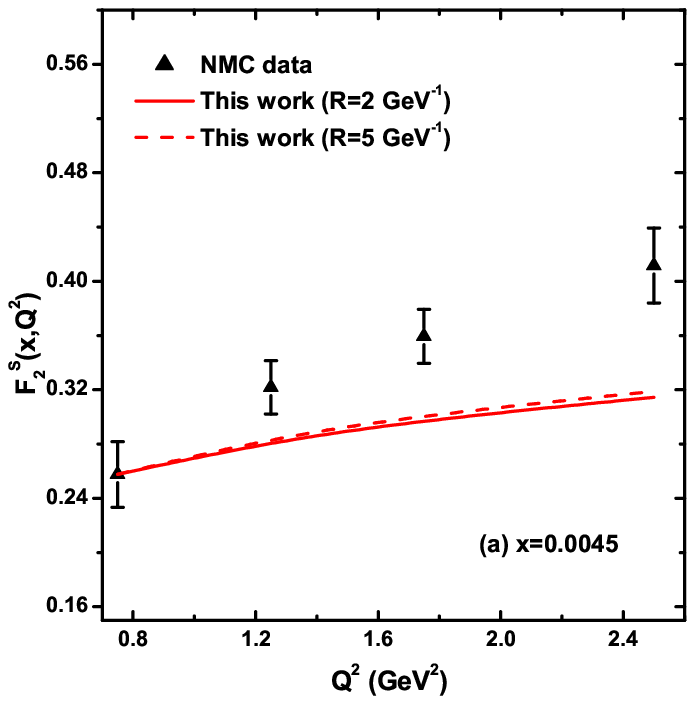}
\includegraphics[width=2.6in, height=2.7in]{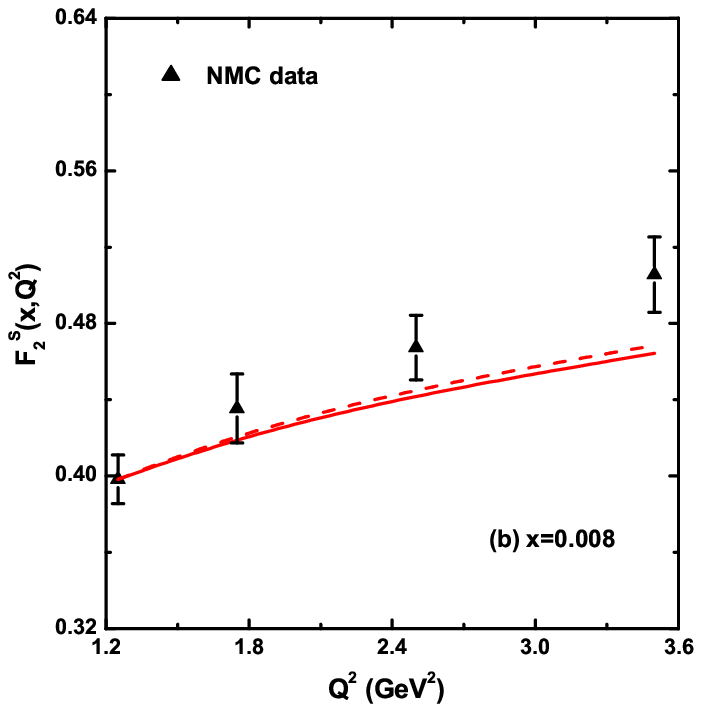}
\caption{\footnotesize {$Q^2$ dependence of singlet structure function $F_2^S(x,Q^2)$ with shadowing corrections at $R=2$ GeV$^{-1}$ (solid curves) and $R=5$ GeV$^{-1}$ (dash curves) computed from Eq. (16) for two representative $x$, viz. $x=0.0045$ and $0.008$  respectively compared to NMC data [23].}}
\label{fig:1}
\end{figure}

\begin{figure}[!htb]
\centering
\includegraphics[width=2.6in, height=2.7in]{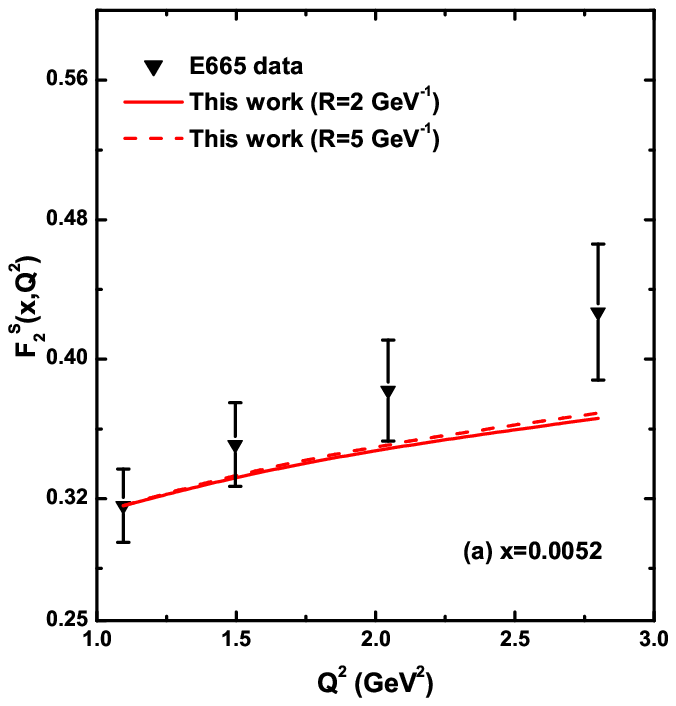}
\includegraphics[width=2.6in, height=2.7in]{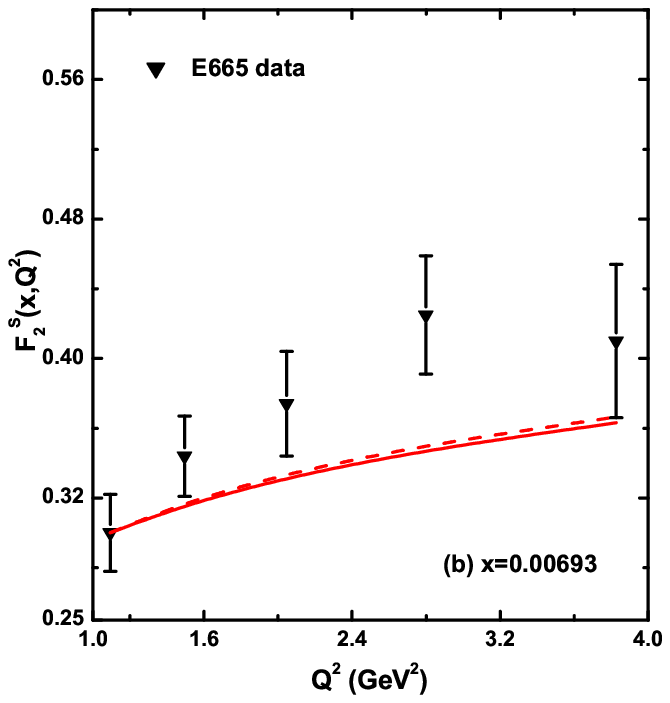}
\includegraphics[width=2.6in, height=2.7in]{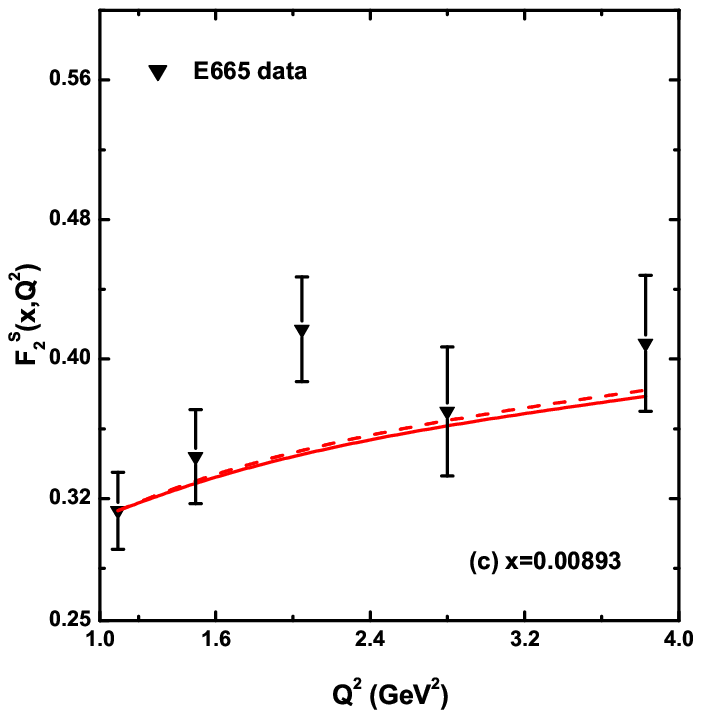}
\caption{\footnotesize {$Q^2$ dependence of singlet structure function $F_2^S(x,Q^2)$ with shadowing corrections at $R=2$ GeV$^{-1}$ (solid curves) and $R=5$ GeV$^{-1}$ (dash curves) respectively computed from Eq. (18) for three fixed $x=0.0052, 0.00693$ and $0.00893$ compared to E665 data [24].}}
\label{fig:1}
\end{figure}

\begin{figure}[!htb]
\centering
\includegraphics[width=2.6in, height=2.7in]{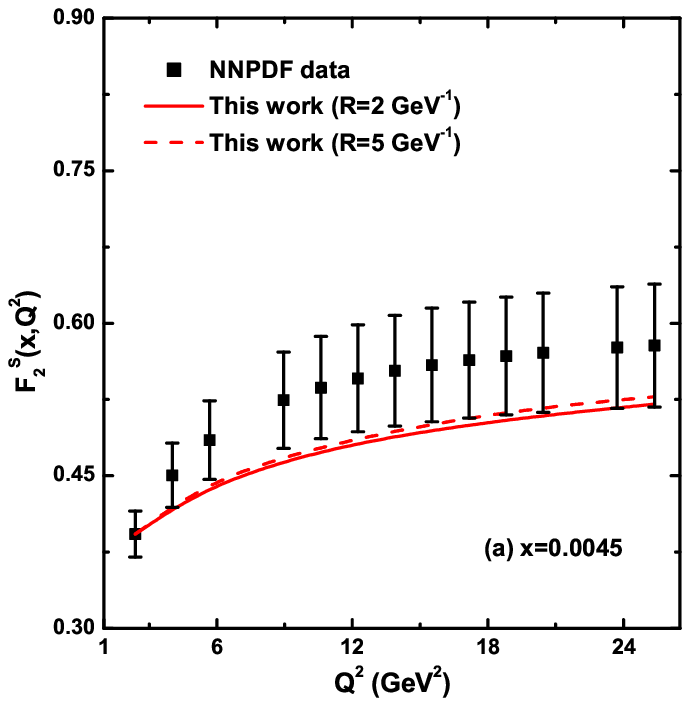}
\includegraphics[width=2.6in, height=2.7in]{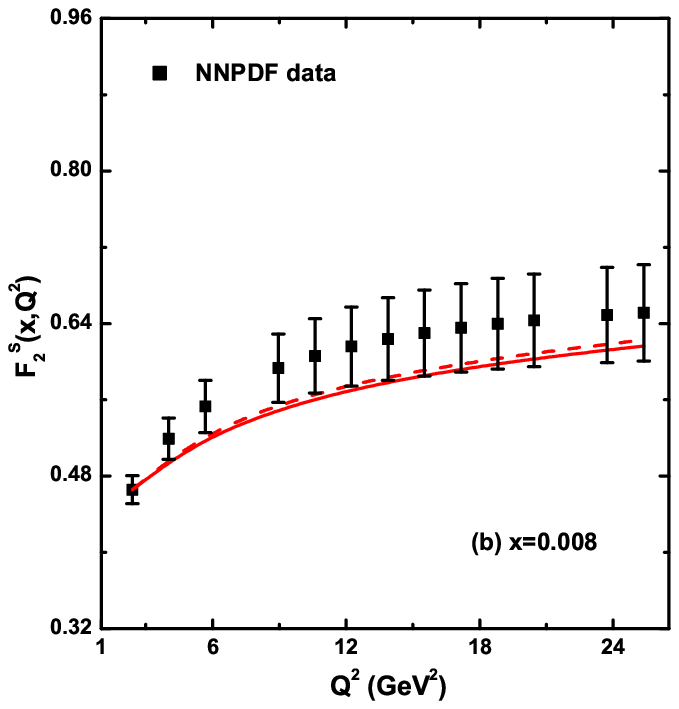}
\caption{\footnotesize {$Q^2$ dependence of singlet structure function $F_2^S(x,Q^2)$ with shadowing corrections at $R=2$ GeV$^{-1}$ (solid curves) and $R=5$ GeV$^{-1}$ (dash curves) respectively computed from Eq. (18) for two fixed $x=0.0045$ and $0.008$ compared to NNPDF data [25].}}
\label{fig:1}
\end{figure}

\begin{figure}[!htb]
\centering
\includegraphics[width=2.6in, height=2.7in]{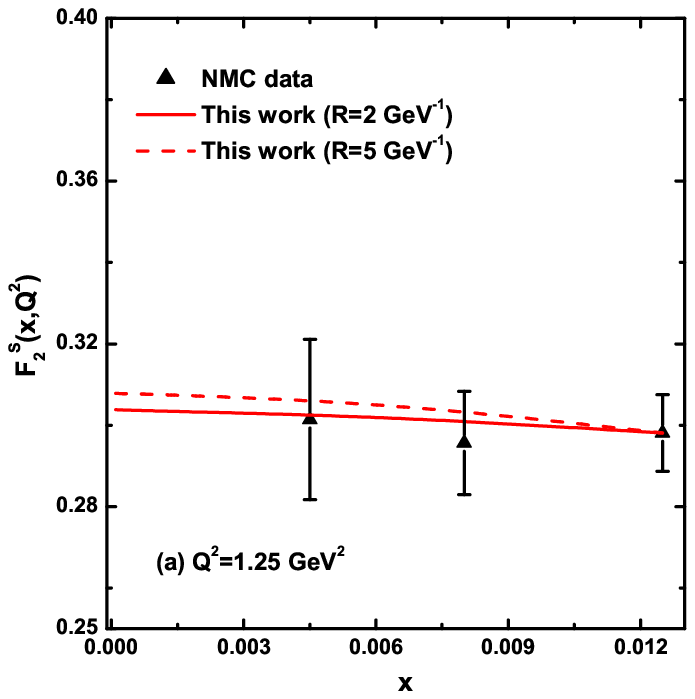}
\includegraphics[width=2.6in, height=2.7in]{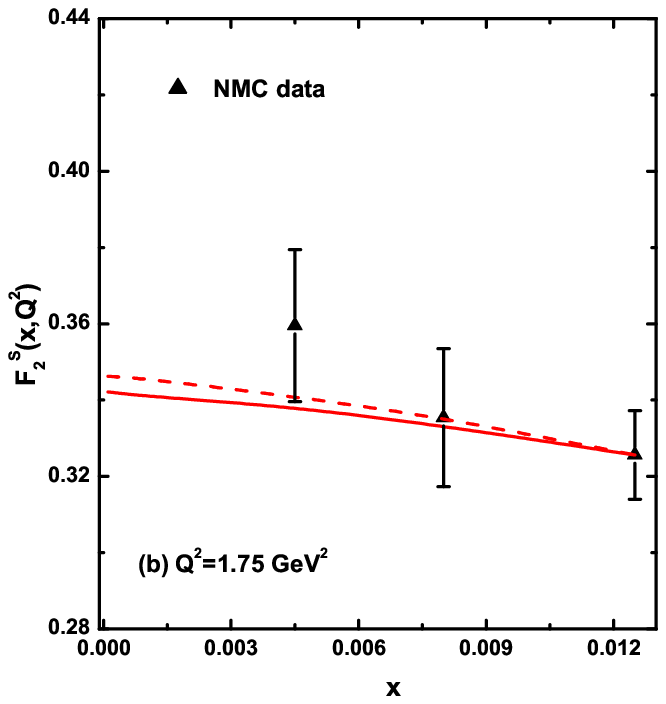}
\includegraphics[width=2.6in, height=2.7in]{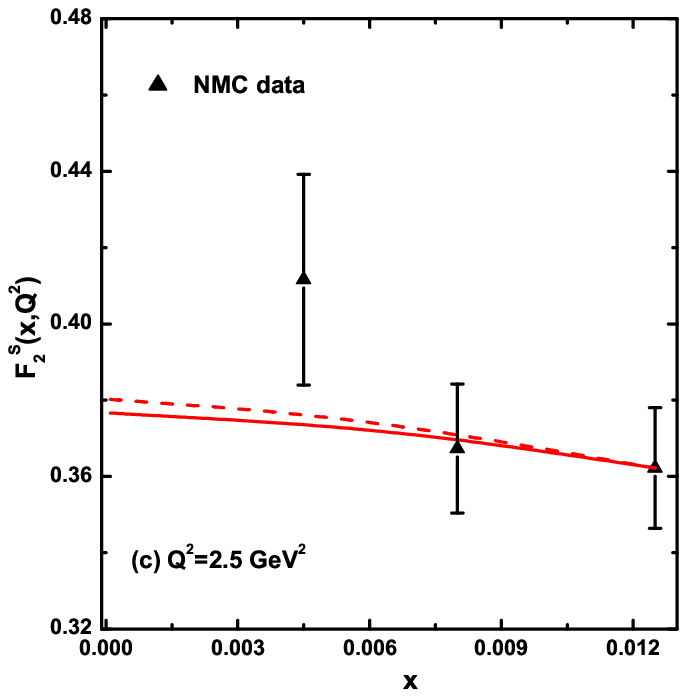}
\caption{\footnotesize {Small-$x$ behavior of singlet structure function $F_2^S(x,Q^2)$ with shadowing corrections at $R=2$ GeV$^{-1}$ (solid curves) and $R=5$ GeV$^{-1}$ (dash curves) respectively computed from Eq.(19) for three representative $Q^2=1.25, 1.75$ and $2.5$ GeV$^2$ compared to NMC data [23].}}
\label{fig:1}
\end{figure}

\begin{figure}[!htb]
\centering
\includegraphics[width=2.6in, height=2.7in]{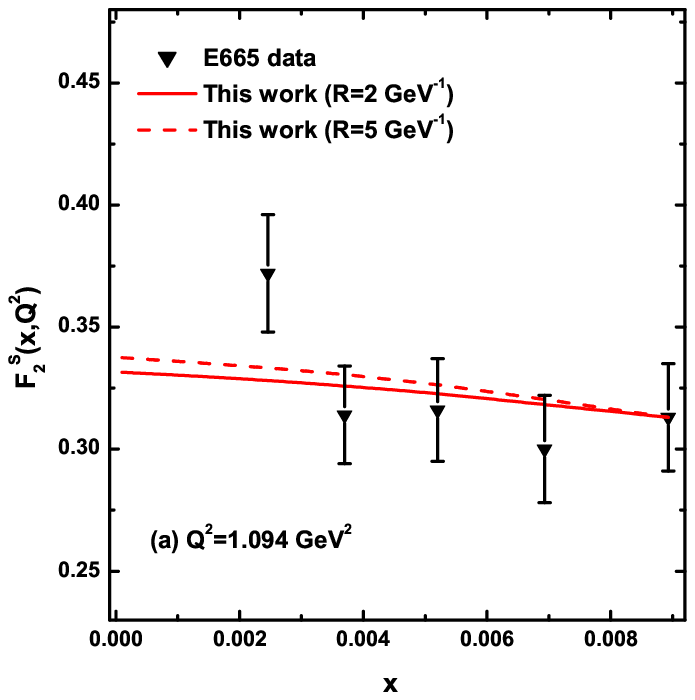}
\includegraphics[width=2.6in, height=2.7in]{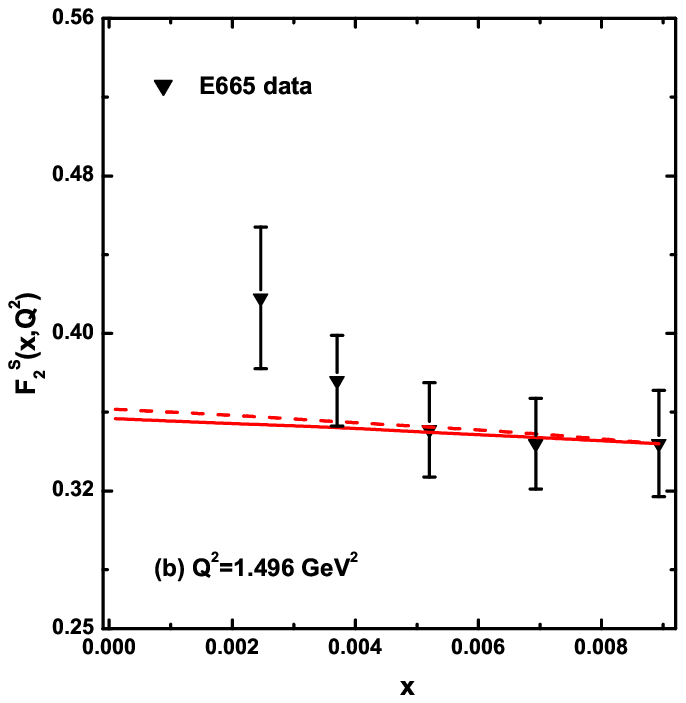}
\includegraphics[width=2.6in, height=2.7in]{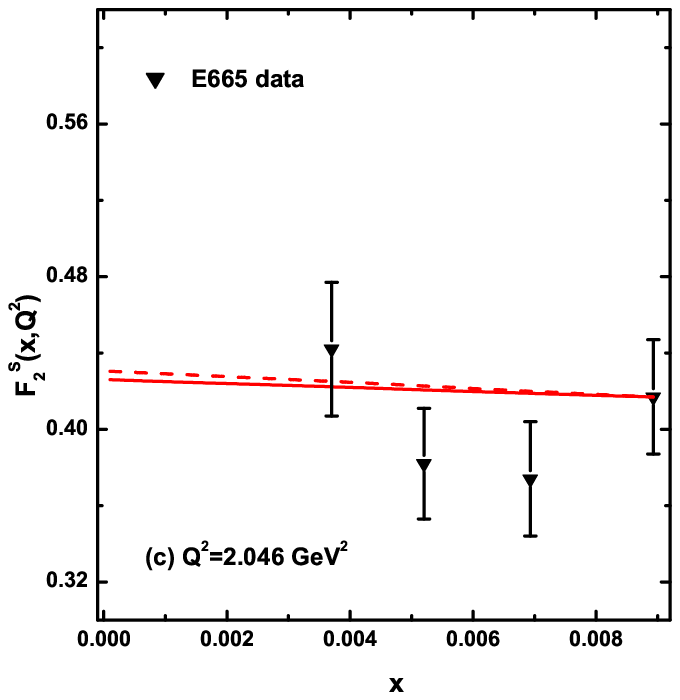}
\includegraphics[width=2.6in, height=2.7in]{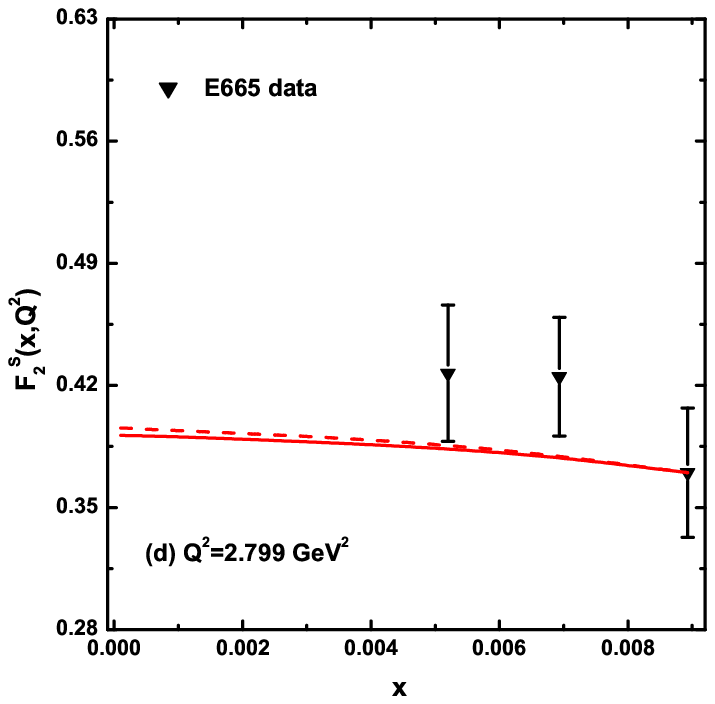}
\caption{\footnotesize {Small-$x$ behavior of singlet structure function $F_2^S(x,Q^2)$ with shadowing corrections at $R=2$ GeV$^{-1}$ (solid curves) and $R=5$ GeV$^{-1}$ (dash curves) respectively computed from Eq.(19) for $Q^2=1.094, 1.496, 2.046$ and $2.799$ GeV$^2$ compared to E665 data [24].}}
\label{fig:1}
\end{figure}

\begin{figure}[!htb]
\centering
\includegraphics[width=2.6in, height=2.7in]{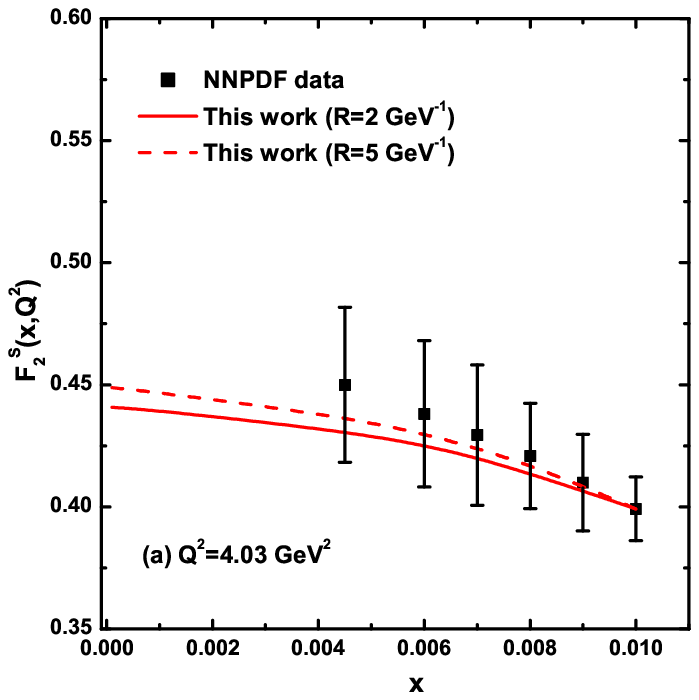}
\includegraphics[width=2.6in, height=2.7in]{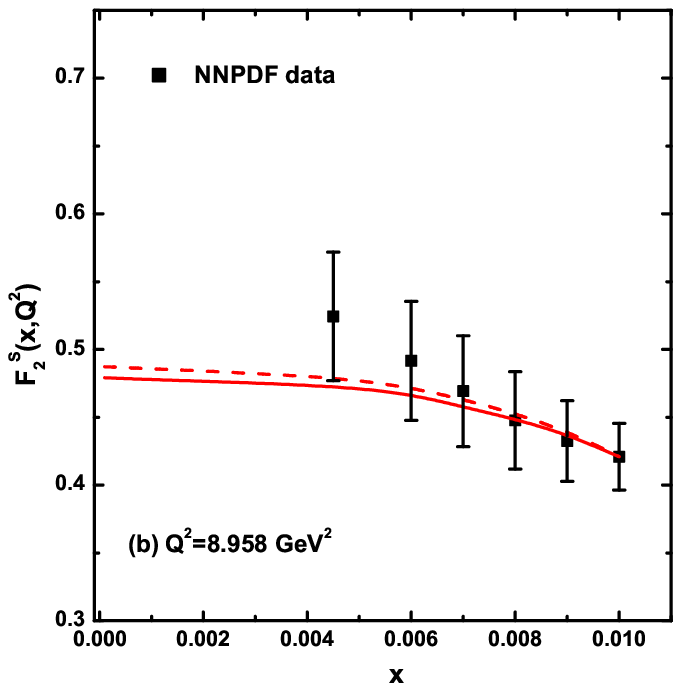}
\includegraphics[width=2.6in, height=2.7in]{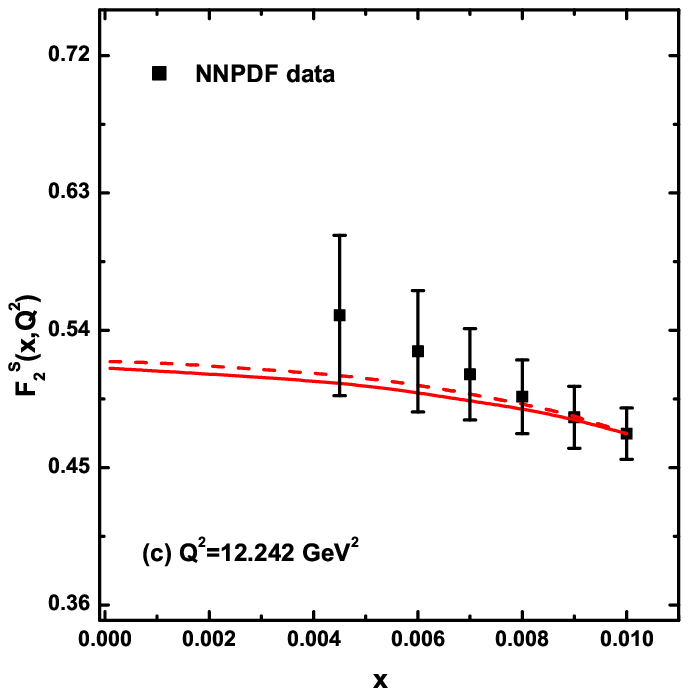}
\includegraphics[width=2.6in, height=2.7in]{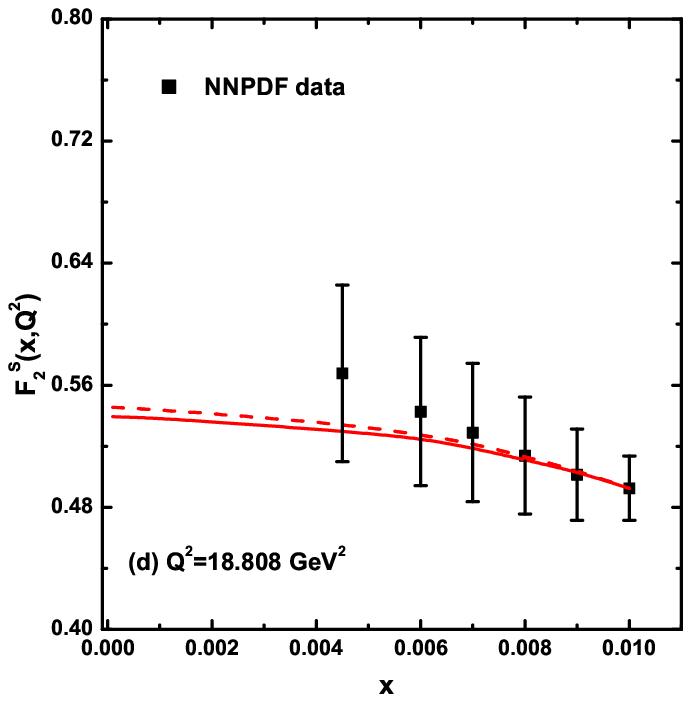}
\caption{\footnotesize {Small-$x$ behavior of singlet structure function $F_2^S(x,Q^2)$ with shadowing corrections at $R=2$ GeV$^{-1}$ (solid curves) and $R=5$ GeV$^{-1}$ (dash curves) respectively computed from Eq.(19) for four fixed $Q^2=4.03, 8.958, 12.242$ and $18.808$ GeV$^2$ compared to NNPDF data [25].}}
\label{fig:1}
\end{figure}
In Fig.1, Fig.2 and Fig.3 we plot the $Q^2$ dependence of singlet structure function $F_2^S(x,Q^2)$ with shadowing corrections computed from Eq. (16) for $R=2$ GeV$^{-1}$ and $R=5$ GeV$^{-1}$ at some representative $x$ and check the compatibility of our predictions with the NMC [23] and E665 [24] experimental data as well as with those obtained by the NNPDF collaboration [25] respectively. It is worthwhile to mention here that the NMC and E665 experiments measured the deuteron structure function $F_2^D$ from which $F_2^S$ can be extracted using the relation $F_2^D=\frac{5}{9}F_2^S$. 
On the other hand, Fig.4, Fig.5 and Fig.6 represent the small-$x$ behavior of $F_2^S(x,Q^2)$ with shadowing corrections computed from Eq. (17) for $R=2$ GeV$^{-1}$ and $R=5$ GeV$^{-1}$ respectively at some fixed values of $Q^2$. 
The consistencies of our results of $x$ dependence of $F_2^S(x,Q^2)$ are also examined with the NMC and E665 experimental data as well as with the NNPDF collaboration respectively.
\begin{figure}[!htb]
\centering
\includegraphics[width=2.8in, height=2.8in]{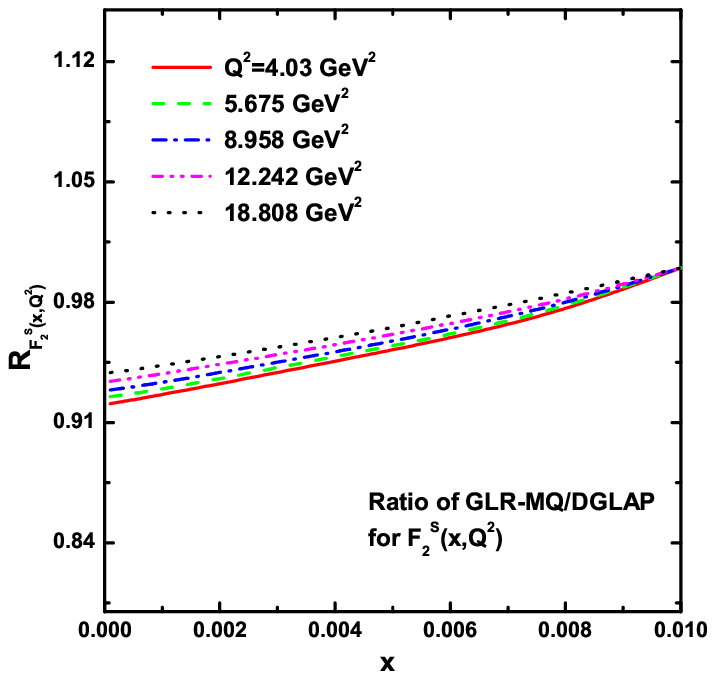}
\caption{\footnotesize {A plot of the ratio of $F_2^S(x,Q^2)$ with shadowing to that without shadowing computed from Eq. (25) for six different bins in $Q^2=4.03, 5.675, 8.958, 12.242$ and $18.808$ GeV$^2$.}}
\label{fig:2}
\end{figure}

\ In the present phenomenological analysis of the shadowing corrections in the evolution of singlet structure functions we assume the function $K(x)$ to be a constant parameter $K$ as a simplest assumption and note that the best fit results are obtained in the range $0.28<K<1.2$ for the entire domain of $x$ and $Q^2$ under study. The computed values of singlet structure functions with shadowing corrections are found to be quite compatible with NMC data in the range ${0.6<{Q^2}<3.6}$ GeV$^2$ and $10^{-4}<{x}<{0.013}$ for $0.52<K<0.9$, E665 data in the range ${1<{Q^2}<4}$ GeV$^2$ and $10^{-4}<{x}<{0.01}$ for $0.28<K<0.86$ and with the NNPDF parametrization in the range ${1<{Q^2}<27}$ GeV$^2$ and $10^{-4}<{x}<{0.011}$ for $0.72<K<1.2$ respectively.

\ We note that the obtained results of singlet structure function complement the perturbative QCD fits at small-$x$, but this attitude is tamed with respect to the nonlinear terms in GLR-MQ equation. We perform our analysis in the kinematic region $0.6\leq{Q^2}\leq{27}$ GeV$^2$ and $10^{-4}\leq{x}\leq{10^{-1}}$ where our predictions of $F_2^S(x,Q^2)$ are found to be legitimate. The effect of shadowing corrections as a consequence of gluon recombination processes in our predictions is observed to be very high at the hot-spot with $R=2$ GeV$^{-1}$ when the gluons are centered within the proton, compared to at $R=5$ GeV$^{-1}$ when the gluons are disseminated throughout the entire proton.

\ Moreover, to examine the effect of nonlinear or shadowing corrections to the singlet structure function in our prediction, we plot the ratio $R_{F_2^S(x,Q^2)}$ of the function $F_2^S(x,t)$ in Fig.5  obtained from the solution of nonlinear GLR-MQ equation to that obtained from the solution of linear DGLAP equation using Eq. (23). We plot the ratio $R_{F_2^S(x,Q^2)}$ as a function of the variable $x$ in the range $10^{-4}\leq{x}\leq{10^{-2}}$ for six representative values $Q^2 =4.03, 5.675, 8.958, 12.242$ and $18.808$ GeV$^2$ respectively. We observe that as $x$ grows smaller the GLR-MQ/DGLAP ratios for ${F_2^S(x,Q^2)}$ decrease which implies that the effect of nonlinearity increases towards small-$x$ due to gloun recombination. We also observe that towards smaller values of $Q^2$ the value of the ratio goes smaller.

\section{Conclusion}

\ To summaries, we solve the nonlinear GLR-MQ equation for sea quark distribution function in leading twist approximation incorporating the well known Regge like ansatz and investigate the effect of nonlinear or shadowing corrections arises due to the gluon recombination processes on the behavior of singlet structure function at small-$x$ and moderate-$Q^2$. We note that the solution of the GLR–MQ equation for singlet structure function with shadowing corrections suggested in this work is found to be valid only in the kinematic domain $10^{-4}<x<10^{-1}$ and $0.4<Q^2<30$ GeV$^2$, where the gluon recombination processes play an important role on the linear QCD evlution. 
Although the obtained results of $F_2^S(x,Q^2)$ increase with increasing $Q^2$ and decreasing $x$, which is in accordance with perturbative QCD fits at small-$x$, however it is very interesting to observe signatures of gluon recombination in our predictions. 
With the inclusion of the nonlinear terms to the linear QCD evolution due to gluon recombination in the small-$x$ region, where density of gluons becomes very high, the steep behaviour of singlet structure function $F_2^S(x,Q^2)$ is slowed down towards small-$x$ leading to a restoration of the Froissart bound. Moreover the effect of shadowing corrections on the behaviour of $F_2^S(x,Q^2)$ with decreasing $x$ become significant at the hot spot with $R=2$ GeV$^{-1}$ when the gluons and the sea quarks are assumed to condensed in a small region within the proton. It is clearly observed that The difference between the results at $R=2$ GeV$^{-1}$ and at $R=5$ GeV$^{-1}$ increase towards smaller vales of $Q^2$ as $x$ decreases. The predictions of the GLR-MQ/DGLAP ratio for ${F_2^S(x,Q^2)}$ also indicate that the gluon recombination processes become significant towards smaller values of $x$ and $Q^2$ .



\begin{thebibliography}{00}
\bibitem{1} A. M. Stasto, K. J. Golec-Biernat, J. Kwiecinski, Phys. Rev. Lett. 86 (2001) 596.
\bibitem{2} J. L. Albacete, C. Marquet, Phys. Rev. Lett. 105 (2010) 162301.
\bibitem{3} A. Dumitru, K. Dusling, F. Gelis, J. Jalilian-Marian, T. Lappi, and R. Venugopalan, Phys. Lett. B 697 (2011) 21.
\bibitem{4} Y.L. Dokshitzer, Sov. Phys. JETP 46 (1977) 641.
\bibitem{5} G. Altarelli, G. Parisi, Nucl. Phys. B 126 (1977) 298.
\bibitem{6} V.N. Gribov, L.N. Lipatov, Sov. J. Nucl. Phys. 15 (1972) 438.
\bibitem{23} M. Froissart, Phys. Rev. 123 (1961) 1053;\\
A. Martin, Phys. Rev. 129 (1963) 1432.
\bibitem{8} L.V. Gribov, E.M. Levin, M.G. Ryskin, Phys. Rep. 100 (1983) 1.
\bibitem{9} A.H. Mueller, J. Qiu, Nucl. Phys. B 268 (1986) 427.
\bibitem{10} A.H. Mueller, Nucl. Phys. B 335 (1990) 115.
\bibitem{11} J. Bartels, E. Levin, Nucl. Phys. B 387 (1992) 617.
\bibitem{12} K. Prytz, Eur. Phys. J. C 22 (2001) 317.
\bibitem{13} K.J. Eskola, H. Honkanen, V.J. Kolhinen, J. Qiu, C.A. Salgado, Nucl. Phys. B 660 (2003) 211.
\bibitem{14} G. Watt, A.D. Martin, M.G. Ryskin, Phys. Lett. B 627 (2005) 97.
\bibitem{15} G. R. Boroun, Eur. Phys. J. A 42 (2009) 251.
\bibitem{16} B. Rezaei, G.R. Boroun, Phys. Lett. B 692 (2010) 247.
\bibitem{17} G. R. Boroun, S. Zarrin, Eur. Phys. J. Plus 128 (2013) 119.
\bibitem{18} M. Devee, J.K. Sarma, Eur. Phys. J. C 74 (2014) 2751.
\bibitem{19} M. Devee, J.K. Sarma, Nucl. Phys. B 885 (2014) 571.
\bibitem{20} P. Phukan, M. Lalung, J. K. Sarma, Nucl. Phys. A 968 (2017) 275.
\bibitem{21} M. Lalung, P. Phukan, J. K. Sarma, Int. J. of Theor. Phys. 56 (2017) 3625.
\bibitem{22} M. Lalung, P. Phukan, J. K. Sarma, arXiv:1801.06360v2 [hep-ph].
\bibitem{23} M. Arneodo et al. (NMC), Nucl. Phys. B 483 (1997) 3.
\bibitem{24} M. R. Adams et al. (E665), Phys. Rev. D 54 (1996) 3006.
\bibitem{25} S. Forte et al. (NNPDF collab.), JHEP 05 (2002) 062.
\bibitem{26} E. Laenen, E. Levin, Nucl. Phys. B 451 (1995) 207.
\bibitem{27} E. Laenen, E. Levin, Annu. Rev. Nucl.Part. Sci. 44 (1994) 199.
\bibitem{28} K. Prytz, Eur. Phys. J. C 22 (2001) 317.
\bibitem{29} E. M. Levin, M. G. Ryskin, Nucl. Phys. B (Proc. Suppl.) 18 (1991) 92.
\bibitem{30} J. Bartels, A. De Roeck, M. Loewe, Z.Phys. C 54 (1992) 635.
\bibitem{31} L.F. Abott, W.B. Atwood, R.M. Barnett, Pys. Rev. D 22 (1980) 582.
\bibitem{32} P.D. Collins, An Introduction to Regge Theory and High-Energy Physics (Cambridge University Press, Cambridge) (1997).
\bibitem{33} A. Donnachie, P.V. Landshoff, Phys. Lett. B 296 (1992) 227.
\bibitem{34} A. Donnachie, P.V. Landshoff, Phys. Lett. B 437 (1998) 408.
\bibitem{35} J. Soffer, O. V. Teryaev, PhysRev D. 56 (1997) 1549.
\bibitem{36} R. Baishya, J.K. Sarma, Phys. Rev. D 74 (2006) 107702.
\bibitem{37} M. Devee, R. Baishya, J.K. Sarma, Eur. Phys. J. C 72 (2012) 2036.
\bibitem{38} A.D. Martin, W.J. Stirling, R.S. Thorne, G. Watt, Eur. Phys. J. C 63 (2009) 189.
\end{thebibliography}
\end{document}